\documentclass[twocolumn,showkeys]{revtex4}
\usepackage[utf8]{inputenc}
\usepackage{graphicx}

\begin{document}

\title{Microscopic model for relativistic hydrodynamics of ideal plasmas}

\author{Pavel A. Andreev}
\email{andreevpa@physics.msu.ru}
\affiliation{Department of General Physics, Faculty of physics, Lomonosov Moscow State University, Moscow, Russian Federation, 119991.}
\affiliation{Faculty of physics, mathematics and natural sciences, Peoples Friendship University of Russia (RUDN University), 6 Miklukho-Maklaya Street, Moscow, 117198, Russian Federation}

\date{\today}

\begin{abstract}
Relativistic hydrodynamics of classic plasmas is derived from the microscopic model in the limit of ideal plasmas.
The chain of equations is constructed step by step starting from the concentration evolution.
It happens that the energy density and the momentum density do not appear at such approach,
but new relativistic-hydrodynamic variables appear in the model.
These variables has no nonrelativistic analogs, but they are reduced to the concentration, the particle current, the pressure (the flux of the particle current) if relativistic effects are dropped.
These variables are reduced to functions of the concentration, the particle current, the pressure
if the thermal velocities are dropped in compare with the relativistic velocity field.
Final equations are presented in the monopole limit of the meanfield (the selfconsistent field) approximation.
Hence, the contributions of the electric dipole moment, magnetic dipole moment, electric quadrupole moment, etc
of the macroscopically infinitesimal element of volume appearing in derived equations are dropped.
\end{abstract}


\maketitle




\section{Introduction}

The relativistic plasmas is found in the astrophysical objects and in the laboratory experiments on the laser-surface interaction and thermonuclear devises.
The role of the relativistic effects is also important for the quantum plasmas
\cite{Hakim book Rel Stat Phys},
\cite{Mahajan PoP 2002},
\cite{Shatashvili ASS 97},
\cite{Shatashvili PoP 99},
\cite{Shatashvili PoP 20},
\cite{Weinberg Gr 72},
\cite{Mahajan PoP 2011},
\cite{Comisso PRL 14},
\cite{Mahajan PRL 03},
\cite{Mahajan PRL 08}.
The degenerate electrons demonstrates the noticeable quantum effects in the relativistic regime
\cite{Andreev PRE 15 SEAW}, \cite{Zhu PRE 10},
\cite{Andreev JPP 21}, \cite{Shukla UFN 10}.
Therefore, the full relativistic model requires the quantum effects as well.
However, this paper is focused on the classical plasmas and the relativistic temperature effects.

The hydrodynamic description of relativistic plasmas with the relativistic temperatures
(so at least one species has relativistic temperature)
is under consideration,
so the temperature (in the energy units) of electrons (like the lightest particle in the plasma)
is comparable with the rest energy of the electron $m_{e}c^{2}$,
where $m_{e}$ is the mass of electron, and $c$ is the speed of light.
One of simplest derivations of the relativistic hydrodynamics of the ideal liquid starts  from the energy-momentum tensor in the rest frame
\cite{Landau v2}, \cite{Landau v6}:
$T^{\alpha\beta}=(\epsilon+\tilde{p})u^{\alpha}u^{\beta}-\tilde{p} g^{\alpha\beta}$,
where the energy density $\epsilon$,
the isotropic pressure $\tilde{p}$ are combined into the enthalpy $w=\epsilon+\tilde{p}$,
and the following components of the four-velocity field in the rest frame $u^{0}=1$, $u^{i}=0$.
Here and below the Greek letters denote the four dimensions $\alpha=0$, $1$, $2$, $3$,
and the Latin letters denote three dimensional (spatial) indexes $i=1$, $2$, $3$.

Knowledge of the energy-momentum tensor leads to the following equation of motion \cite{Landau v6}
\begin{equation}\label{RHD2018Cl d T a b 0} \partial_{\beta}T^{\alpha\beta}=0.\end{equation}
If we consider the charged particles interacting with the electromagnetic field
we need to extend equation (\ref{RHD2018Cl d T a b 0})
\begin{equation}\label{RHD2018Cl d T a b F} \partial_{\beta}T^{\alpha\beta}=q n F^{\alpha\beta}u^{\beta},\end{equation}
where $q$ is the charge of particle,
$n$ is the concentration,
and
$F^{\alpha\beta}$ is the tensor of electromagnetic field.

Here the rest frame is defined for the macroscopically motionless fluid.
However, the rest frame can be defined in the presence of the local macroscopic flows
or/and turbulence.

Next step is generalization of the presented in the rest frame hydrodynamic equations to the arbitrary inertial frame.
It includes the Lorentz transformation for the representation of hydrodynamic equations
(or the energy-momentum tensor) in the arbitrary frame,
where the arbitrary frame moves with the velocity $u^{i}$ relatively the rest frame.

Apparently, the velocity of the frame $u^{i}$ is not related to the local flows
which can exist in the fluid and described by the velocity field.
Moreover, it is essential to point out that
velocity $u^{i}$ is a constant.
It does not depend on the position
while the velocity field of the fluid is a function of space.
The following remark is in order.
The application of the described method locally cannot be done,
since the Lorentz transformation is a global transformation.
Described transformation is useful for the study of beams or global flows \emph{with nonrelativistic temperatures}.
It can be assumed that
the energy density and the pressure in the energy-momentum tensor correspond to the relativistic temperatures via suitable equations of state.
Since the Lorentz transformation makes the motion of the whole system or the motion of the observer relatively medium and does not related to the appearance of the local flows.

The energy-momentum tensor $T^{\alpha\beta}$ is constructed of all possible tensors and vectors in the isotropic fluid
which are the Kronecker symbol $\delta^{\alpha\beta}$ and the velocity vector $v^{\alpha}$.
However, the presented below analysis of the relativistically hot plasmas demonstrates presence of another nontrivial four vector.
Hence, the energy-momentum tensor will be reconsidered after the introduction of our model.

The comments described above show some problems of the hydrodynamic model constructed phenomenologically on the macroscopic scale.
Anyway, present level of knowledge requires some microscopic justification of the macroscopic models.
Sometimes authors refer to the kinetic theory as the microscopic model.
Here, we mean that
the microscopic level is the scale,
where motion of each particle is distinguishable
(the scale of electrons and protons for the hydrogen plasmas).
From this point of view the kinetics is the macroscopic method of description formulated in the six dimensional space of coordinates and momentums.
In our derivation we avoid derivation of the kinetic model as an intermediate stage.
We directly derive the hydrodynamics from the microscopic motion of particles.

This paper is organized as follows.
In Sec. II derivation of the relativistic hydrodynamic model based on exact microscopic motion of particles is shown.
The basic definitions and method of derivation are demonstrated.
In Sec. III the suggested relativistic hydrodynamic model is represented in terms of the velocity field.
In Sec. IV method of derivation of equations of state is described to make truncation of the set of equations.
In Sec. V necessity of the covariant structure of hydrodynamic equations.
In Sec. VI the microscopic structure of the energy-momentum tensor appearing via the microscopic derivation is demonstrated and discussed.
In Sec. VII the spectrum of high-frequency longitudinal plane waves in the relativistically hot isotropic plasmas.
In Sec. VIII the zero-temperature limit of the model is found to demonstrate its agreement with well-known results
for the relativistic beams in the low temperature plasmas.
In Sec. IX brief discussion of the plane waves in the relativistically hot magnetized plasmas is given.
In Sec. X a brief summary of obtained results is presented.


\section{Model}

Our goal is the derivation of the macroscopic hydrodynamic equations for the relativistic plasmas
with the relativistically large temperatures.
We are going to trace exact microscopic evolution of particles obeying the relativistic modification of the Newton equations of motion.


\subsection{Basic definitions}

We start our derivation with the definition of the concentration of particles.
In classical physics the particle is modeled as the point-like object.
Hence, its mathematical representation is the delta-function
(one particle in the zero volume space gives infinite particle number density in one point and zero density in other point).
Therefore, the microscopic concentration of the classical system of particles is the sum of delta functions:
\begin{equation}\label{RHD2018Cl concentration mic}
n_{mic}(\textbf{r},t)=\sum_{i=1}^{N}\delta(\textbf{r}-\textbf{r}_{i}(t)),\end{equation}
where the subindex \textit{mic} refers to the microscopic definition of the concentration of particles.
Function $\textbf{r}_{i}(t)$ is the radius vector of $i$-th particle.
Its evolution happens in accordance with the Newton equation of motion,
where the interaction with all surrounding particles is included.
Presented here microscopic definition of concentration evolve with accordance with exact microscopic evolution of particles.
So, if one wants to find its evolution one need to solve set of Newton equations of motion.
Our goal is to construct approximate macroscopic model to capture main features of the relativistic plasmas.

However, if we consider the macroscopic theory
we need to introduce the macroscopically infinitesimal element of volume $\Delta$
and present the number of particles in this element of volume
\begin{equation}\label{RHD2018Cl concentration definition}
n(\textbf{r},t)\equiv n_{mac}(\textbf{r},t)=\frac{1}{\Delta}\int_{\Delta}d\mbox{\boldmath $\xi$}
\sum_{i=1}^{N}\delta(\textbf{r}+\mbox{\boldmath $\xi$}-\textbf{r}_{i}(t)). \end{equation}
The presented integral does not directly related to the simplified point-like structure of particles.
It counts number of particles in the chosen vicinity $\Delta$ of the chosen point $\textbf{r}$
since the integral of the delta function is equal to one if the particle is in the vicinity
and the integral of the delta function is equal to zero if the particle is outside of the vicinity.
Equation (\ref{RHD2018Cl concentration definition}) can be used to the center of mass of finite size objects.
However, for the point-like objects,
equation (\ref{RHD2018Cl concentration definition}) is equivalent to microscopic concentration (\ref{RHD2018Cl concentration mic}).

The macroscopic concentration can be rewritten in different form
which can provide more physical inside of the presented definition (\ref{RHD2018Cl concentration definition}):
\begin{equation}\label{RHD2018Cl concentration REdefinition}
n(\textbf{r},t)=\frac{1}{m\Delta} \sum_{i=1}^{N(\textbf{r},t)}m_{i}, \end{equation}
where all $m_{i}$ are equal to each other
(the concentration is defined for each species in plasmas).
Function $N(\textbf{r},t)$ is the number of particles in the $\Delta$-vicinity of point $\textbf{r}$ at the fixed moment in time $t$.
These definitions (\ref{RHD2018Cl concentration definition}) and (\ref{RHD2018Cl concentration REdefinition}) are equivalent to each other,
but (\ref{RHD2018Cl concentration REdefinition}) less useful for the derivation of the continuity equation
since the unknown function $N(\textbf{r},t)$ is present in the upper limit of summation.
The integral operator in equation (\ref{RHD2018Cl concentration definition})
\begin{equation}\label{RHD2018Cl concentration definition}
\int_{\Delta}d\mbox{\boldmath $\xi$}
\sum_{i=1}^{N}\delta(\textbf{r}+\mbox{\boldmath $\xi$}-\textbf{r}_{i}(t))=N(\textbf{r},t) \end{equation}
counts the number of particles $N(\textbf{r},t)$ following their exact motion in accordance with the trajectory of each particle $\textbf{r}_{i}(t)$.
Equation (\ref{RHD2018Cl concentration REdefinition}) can be rewritten with no application of mass of particles $m_{i}$
via the count of units:
$n(\textbf{r},t)=\sum_{i=1}^{N(\textbf{r},t)}1_{i}/\Delta=N(\textbf{r},t)/\Delta$.

At initial step we have no conditions on volume $\Delta$.
If we need to present a macroscopic theory
we should have $\Delta$ large enough
so $\Delta$-vicinity of each point of space contains macroscopically large number of particles.
So, $\Delta$-volume can play the role of the macroscopically infinitesimal element of space.
On the other hand we can show transition from equation (\ref{RHD2018Cl concentration definition})
to equation (\ref{RHD2018Cl concentration mic}) at $\Delta\rightarrow0$.
If the point of space $\textbf{r}$ contains a particle (or several point-like particles)
we have $n(\textbf{r},t)=1/(\Delta\rightarrow0)\rightarrow\infty$.
If the point of space $\textbf{r}$ does not contain any particle
we find $n(\textbf{r},t)=lim_{\Delta\rightarrow0}(0/\Delta)=0$.
So, we have $N$ delta functions giving $N$ infinite values at points $\textbf{r}=\textbf{r}_{i}(t)$ for $i\in[1,N]$
and zero values in other points.
So, we have same distribution as in equation (\ref{RHD2018Cl concentration mic}).

A possible candidate for the concentration definition is
\begin{equation}\label{RHD2018Cl concentration a possible definition} \tilde{n}(\textbf{r},t)=\frac{1}{\Delta}\int_{\Delta}d\mbox{\boldmath $\xi$}\sum_{i=1}^{N}\gamma_{i}\delta(\textbf{r}+\mbox{\boldmath $\xi$}-\textbf{r}_{i}(t)), \end{equation}
where $\gamma_{i}=(1-\textbf{v}_{i}^{2}(t)/c^{2})^{-1/2}$
with $\textbf{v}_{i}(t)=d\textbf{r}_{i}(t)/dt$ the velocity of $i$th particle.
However, function $\tilde{n}$ is proportional to the energy density $\tilde{n}=\varepsilon(\textbf{r},t)/mc^{2}$,
hence it obviously does not satisfy the continuity equation.

Presented method is the three-dimensional reduction of the method of microscopic derivation of the relativistic kinetics \cite{Kuz'menkov 91}.

\subsection{Continuity equation}

Analysis of the concentration dynamics can be obtained without the equation of motion
since it is defined by the kinematic effects
$$\partial_{t}n(\textbf{r},t)
=\frac{1}{\Delta}\int_{\Delta}d\mbox{\boldmath $\xi$}\sum_{i=1}^{N} \partial_{t}\delta(\textbf{r}+\mbox{\boldmath $\xi$}-\textbf{r}_{i}(t))$$
\begin{equation}\label{RHD2018Cl concentration derivative}
=\frac{1}{\Delta}\int_{\Delta}d\mbox{\boldmath $\xi$}\sum_{i=1}^{N} (-\textbf{v}_{i}(t))
\cdot\partial_{\textbf{r}}\delta(\textbf{r}+\mbox{\boldmath $\xi$}-\textbf{r}_{i}(t)).
\end{equation}
The derivative on the space variable $\partial_{\textbf{r}}$ can be placed outside of the integral in the last term.
Hence, the time evolution of the concentration leads to the appearance of the particles current:
\begin{equation}\label{RHD2018Cl particle current definition}
\textbf{j}(\textbf{r},t)=\frac{1}{\Delta}\int_{\Delta}d\mbox{\boldmath $\xi$}\sum_{i=1}^{N} \textbf{v}_{i}(t)
\delta(\textbf{r}+\mbox{\boldmath $\xi$}-\textbf{r}_{i}(t)).
\end{equation}

Summing up presented results as the well-known continuity equation:
\begin{equation}\label{RHD2018Cl continuity equation} \partial_{t}n+\nabla\cdot\textbf{j}=0. \end{equation}

There is no statistical averaging in the described prescription.
Moreover, it is not necessary to use the statistics
while we trace the microscopic motion itself.
However, some short notations (which can remind statistical physics) are in order.
For example, the particle current appears as
the action of operator
\begin{equation}\label{RHD2018Cl brackets}\langle ... \rangle\equiv
\frac{1}{\Delta}\int_{\Delta}d\mbox{\boldmath $\xi$}\sum_{i=1}^{N} ... \delta(\textbf{r}+\mbox{\boldmath $\xi$}-\textbf{r}_{i}(t))\end{equation}
on the velocities of particles.
Therefore, the particle current can be written in a short form: $\textbf{j}=\langle \textbf{v}_{i}\rangle$.

The nonrelativistic hydrodynamics is considered within this method in Ref. \cite{Drofa TMP 96}.
It is also briefly discussed in Ref. \cite{Andreev PIERS 2012}.

\subsection{Equation for the current evolution}

Derivation of the other equations including the current evolution equation
requires equation of motion for each particle
to trace their exact motion in terms of collective (macroscopic) variables.
we use the relativistic Newton equations in terms of the velocity evolution
(or the acceleration caused by the interaction) \cite{Landau v2} (see section 17): 
\begin{equation}\label{RHD2018Cl Newton eq} \dot{\textbf{v}}_{i}=\frac{e_{i}}{m_{i}}\sqrt{1-\frac{\textbf{v}_{i}^{2}}{c^{2}}} \biggl[\textbf{E}_{i}+\frac{1}{c}[\textbf{v}_{i}\times\textbf{B}_{i}]-\frac{1}{c^{2}}\textbf{v}_{i}(\textbf{v}_{i}\cdot\textbf{E}_{i})\biggr], \end{equation}
where $\textbf{v}_{i}=\textbf{v}_{i}(t)$,
$\textbf{E}_{i}=\textbf{E}(\textbf{r}_{i}(t),t)$ and
$\textbf{B}_{i}=\textbf{B}(\textbf{r}_{i}(t),t)$ are the electric and magnetic fields
in the point $\textbf{r}_{i}(t)$ and at time $t$ acting on $i$th particle.

We consider the evolution of the particle current $\textbf{j}(\textbf{r},t)$.
To this end we calculate the time derivative of the current:
$$\partial_{t}j^{a}(\textbf{r},t)
=\frac{1}{\Delta}\int_{\Delta}d\mbox{\boldmath $\xi$}\sum_{i=1}^{N} v_{i}^{a}(t)\partial_{t}\delta(\textbf{r}+\mbox{\boldmath $\xi$}-\textbf{r}_{i}(t))$$
\begin{equation}\label{RHD2018Cl current derivative}
+\frac{1}{\Delta}\int_{\Delta}d\mbox{\boldmath $\xi$}\sum_{i=1}^{N} \dot{v}_{i}^{a}(t)\delta(\textbf{r}+\mbox{\boldmath $\xi$}-\textbf{r}_{i}(t)).
\end{equation}
We present the presented equation introducing the flux of the particle current $\Pi^{ab}$ and the force field $F^{a}$
\begin{equation}\label{RHD2018Cl current derivative Euler equation}
\partial_{t}j^{a}+\partial_{b}\Pi^{ab}=F^{a},
\end{equation}
where
\begin{equation}\label{RHD2018Cl def Pi}
\Pi^{ab}=\frac{1}{\Delta}\int_{\Delta}d\mbox{\boldmath $\xi$}\sum_{i=1}^{N} v_{i}^{a}(t)v_{i}^{b}(t)\delta(\textbf{r}+\mbox{\boldmath $\xi$}-\textbf{r}_{i}(t)),
\end{equation}
and
\begin{equation}\label{RHD2018Cl force field via acceleration}
F^{a}=\frac{1}{\Delta}\int_{\Delta}d\mbox{\boldmath $\xi$}\sum_{i=1}^{N} \dot{v}_{i}^{a}(t)\delta(\textbf{r}+\mbox{\boldmath $\xi$}-\textbf{r}_{i}(t)).
\end{equation}
Strictly speaking the introduced force field $F^{a}$ is not exactly the force field,
since the force causes the change of momentum $\dot{\textbf{p}}_{i}=\mathcal{F}$.
While $F^{a}$ is the acceleration of particles given by equation (\ref{RHD2018Cl Newton eq})
in terms of the electromagnetic field acting on the particle.

Next, we substitute the acceleration from equation (\ref{RHD2018Cl Newton eq}) in the force field $F^{a}$ (\ref{RHD2018Cl force field via acceleration})
and find the expression containing new hydrodynamic functions
$$F^{a}=\frac{1}{\Delta}\int_{\Delta}d\mbox{\boldmath $\xi$}\sum_{i=1}^{N} \frac{e_{i}}{m_{i}\gamma_{i}}E^{a}(\textbf{r}+\mbox{\boldmath $\xi$},t) \delta(\textbf{r}+\mbox{\boldmath $\xi$}-\textbf{r}_{i}(t))$$
$$+\frac{1}{c}\varepsilon^{abc}\frac{1}{\Delta}\int_{\Delta}d\mbox{\boldmath $\xi$}\sum_{i=1}^{N} \frac{e_{i}}{m_{i}\gamma_{i}}v_{i}^{b}B^{c}(\textbf{r}+\mbox{\boldmath $\xi$},t) \delta(\textbf{r}+\mbox{\boldmath $\xi$}-\textbf{r}_{i}(t))$$
\begin{equation}\label{RHD2018Cl force field via acceleration}
-\frac{1}{c^{2}}\frac{1}{\Delta}\int_{\Delta}d\mbox{\boldmath $\xi$}\sum_{i=1}^{N} \frac{e_{i}}{m_{i}\gamma_{i}}v_{i}^{a}v_{i}^{b}E^{b}(\textbf{r}+\mbox{\boldmath $\xi$},t) \delta(\textbf{r}+\mbox{\boldmath $\xi$}-\textbf{r}_{i}(t)).
\end{equation}
In equation (\ref{RHD2018Cl force field via acceleration})
we apply the replacement of coordinate of particles in the argument of the electromagnetic field $\textbf{r}_{i}(t)$
on $\textbf{r}+\mbox{\boldmath $\xi$}$ using the $\delta$-function $\delta(\textbf{r}+\mbox{\boldmath $\xi$}-\textbf{r}_{i}(t))$.

We consider the monopole approximation of the electric and magnetic fields
$E^{a}(\textbf{r}+\mbox{\boldmath $\xi$},t)\approx E^{a}(\textbf{r},t)$,
and
$B^{a}(\textbf{r}+\mbox{\boldmath $\xi$},t)\approx B^{a}(\textbf{r},t)$.
It corresponds to the mean-field (self-consistent field) approximation traditionally applied at the study of the plasmas.
This approximation is possible if the electric and magnetic fields have small change on the scale of $\Delta$-vicinity.
For instance, if the particle is under the action of the electromagnetic wave
its wavelength $\lambda$ should be large in compare with the radius of the vicinity:
$\lambda\gg\sqrt[3]{\Delta}$.

The mean-field approximation simplify the force field to the following form
$$F^{a}=E^{a}(\textbf{r},t)\cdot\frac{1}{\Delta}\int_{\Delta}d\mbox{\boldmath $\xi$}\sum_{i=1}^{N} \frac{e_{i}}{m_{i}\gamma_{i}} \delta(\textbf{r}+\mbox{\boldmath $\xi$}-\textbf{r}_{i}(t))$$
$$+\frac{1}{c}\varepsilon^{abc}B^{c}(\textbf{r},t)\cdot\frac{1}{\Delta}\int_{\Delta}d\mbox{\boldmath $\xi$}\sum_{i=1}^{N} \frac{e_{i}}{m_{i}\gamma_{i}}v_{i}^{b} \delta(\textbf{r}+\mbox{\boldmath $\xi$}-\textbf{r}_{i}(t))$$
\begin{equation}\label{RHD2018Cl force field expanded}
-\frac{1}{c^{2}}E^{b}(\textbf{r},t)\cdot\frac{1}{\Delta}\int_{\Delta}d\mbox{\boldmath $\xi$}\sum_{i=1}^{N} \frac{e_{i}}{m_{i}\gamma_{i}}v_{i}^{a}v_{i}^{b} \delta(\textbf{r}+\mbox{\boldmath $\xi$}-\textbf{r}_{i}(t)).
\end{equation}
This presigere gives the mean-field (the self-consistent) approximation.

Evolution of the particle current leads to three new hydrodynamic variables
\begin{equation}\label{RHD2018Cl Gamma function definition} \Gamma=\frac{1}{\Delta}\int_{\Delta}d\mbox{\boldmath $\xi$}\sum_{i=1}^{N} \frac{1}{\gamma_{i}} \delta(\textbf{r}+\mbox{\boldmath $\xi$}-\textbf{r}_{i}(t)),\end{equation}
\begin{equation}\label{RHD2018Cl Theta function definition} \Theta^{a}=\frac{1}{\Delta}\int_{\Delta}d\mbox{\boldmath $\xi$}\sum_{i=1}^{N} \frac{1}{\gamma_{i}}v_{i}^{a} \delta(\textbf{r}+\mbox{\boldmath $\xi$}-\textbf{r}_{i}(t)),\end{equation}
and
\begin{equation}\label{RHD2018Cl Xi function definition}
\Xi^{ab}=\frac{1}{\Delta}\int_{\Delta}d\mbox{\boldmath $\xi$}\sum_{i=1}^{N} \frac{1}{\gamma_{i}}v_{i}^{a}v_{i}^{b} \delta(\textbf{r}+\mbox{\boldmath $\xi$}-\textbf{r}_{i}(t)).
\end{equation}

Overall, we find the following equation
\begin{equation}\label{RHD2018Cl current derivative Euler equation explicit}
\partial_{t}j^{a}+\partial_{b}\Pi^{ab}
=\frac{e}{m}\biggl(\Gamma E^{a} +\frac{1}{c}\varepsilon^{abc}\Theta^{a}B^{c} -\frac{1}{c^{2}}\Xi^{ab} E^{b}\biggr),
\end{equation}
which is the particle current evolution equation.
It is one of the generalizations of the nonrelativistic Euler equation for the velocity field on the relativistic regime.

A brief presentation of this model is given in short report \cite{Andreev 2021 05}
Some steps towards present model are made in Ref. \cite{Andreev arXiv 12 relativistic}.

\subsection{New relativistic hydrodynamic variables}

Relativistic equation for the particle current evolution
gives three new hydrodynamic functions $\Gamma$, $\Theta^{a}$, and $\Xi^{ab}$.
These hydrodynamic functions do not exist in the nonrelativistic limit.
The particle current $j^{a}$ coincides with the momentum density in the nonrelativistic limit.
Moreover, the hydrodynamic Gamma function $\Gamma$ tends to the concentration $n$ in the nonrelativistic limit.
The hydrodynamic Theta function $\Theta^{a}$ has same nonrelativistic limit as the current of particles $j^{a}$ and the momentum density.
The hydrodynamic Xi function has same form in the nonrelativistic regime as the flux of the particle current $\Pi^{ab}$ (\ref{RHD2018Cl def Pi})
and the momentum flux (containing the pressure).

We have two different functions in the relativistic hydrodynamics:
the current of particles and the momentum density.
Hence, we need to choose
which of them should be included in the model.
The answer on this question follows from the Maxwell equations,
where the electromagnetic field is caused by the concentration and the current of particles.
Therefore, we consider the momentum density as a function
which is unnecessary for application in our model.

Hydrodynamic Gamma function $\Gamma$ is
the "average" reverse relativistic gamma factor $\Gamma=\langle\frac{1}{\gamma_{i}}\rangle$
(see equation (\ref{RHD2018Cl Gamma function definition})).
Hydrodynamic Theta vector function $\Theta^{a}$ is the current  of the reverse gamma factor.
So, vector field $\Theta^{a}$ is the current of the scalar function $\Gamma$:
$\Theta^{a}=\langle v_{i}^{a}/\gamma_{i}\rangle$ (see equation (\ref{RHD2018Cl Theta function definition})).
Next, the tensor field $\Xi^{ab}$ is the flux or current of the vector field $\Theta^{a}$:
$\Xi^{ab}=\langle v_{i}^{a}v_{i}^{b}/\gamma_{i}\rangle$
(see equation \ref{RHD2018Cl Xi function definition}).

\subsection{Equations evolution for the new relativistic hydrodynamic variables}

For the further development of the relativistic hydrodynamic
equations consider the evolution of the hydrodynamic Gamma and Theta functions.

It is well-known that
the application of nonrelativistic hydrodynamic composed of the continuity and Euler equations
gives incorrect coefficient for the thermal contributions in the spectrum
in compare with the kinetic results \cite{Tokatly PRB 99}, \cite{Tokatly PRB 00}.
However, the extension of the hydrodynamic model including the pressure evolution equation allows to improve the results
\cite{Andreev JPP 21}, \cite{Tokatly PRB 99}, \cite{Tokatly PRB 00}, \cite{Miller PoP 16}.
Here, we try to create a relativistic minimal coupling hydrodynamic model,
which is based on the evolution of the scalar and vector functions
$n$, $j^{a}$, $\Gamma$, $\Theta^{a}$,
where the tensor fields of the higher tensor rank
(like $\Pi^{ab}$ and $\Xi^{ab}$)
are expressed using equations of state.

\subsubsection{Equation of the Gamma function evolution}

We consider the temporal evolution of the Gamma function according to the described method
(by the calculation of the time derivative of the definition of required function)
\begin{equation}\label{RHD2018Cl equation of Gamma evolution 1}
\partial_{t}\Gamma+\partial_{a}\Theta^{a}= -\frac{1}{c^{2}}\frac{1}{\Delta}\int_{\Delta}d\mbox{\boldmath $\xi$}\sum_{i=1}^{N} \frac{1}{\gamma_{i}}v_{i}^{a}\dot{v}_{i}^{a} \delta_{i},
\end{equation}
where
the second term in equation comes from differentiating of the delta function,
the last term appears from differentiating of the reverse gamma factor,
$\delta_{i}\equiv\delta(\textbf{r}+\mbox{\boldmath $\xi$}-\textbf{r}_{i}(t))$
is the short representation of the delta function.
Moreover, let us mention a technical detail
that the gamma factor is in the denominator in the last term of equation (\ref{RHD2018Cl equation of Gamma evolution 1})

We consider the right-hand side of equation (\ref{RHD2018Cl equation of Gamma evolution 1}) using the relativistic Newton equation (\ref{RHD2018Cl Newton eq})
\begin{equation}\label{RHD2018Cl} \biggl\langle\frac{1}{\gamma_{i}}v_{i}^{a}\dot{v}_{i}^{a}\biggr\rangle=\frac{e}{m}\langle v_{i}^{a}E_{i}^{a}\rangle-\frac{1}{c^{2}}\frac{e}{m}\langle v_{i}^{a}\textbf{v}_{i}^{2}E_{i}^{a}\rangle, \end{equation}
where we use that
$\gamma_{i}v_{i}^{a}\dot{v}_{i}^{a}
=\frac{e}{m}v_{i}^{a}(E^{a}-\frac{1}{c^{2}}v_{i}^{a}(\textbf{v}_{i}\cdot\textbf{E}))$
$=\frac{e}{m}(\textbf{v}_{i}\cdot\textbf{E}))(1-\frac{1}{c^{2}}\textbf{v}_{i}^{2})$.
It can be rewritten via the gamma factor
$\frac{e}{m}(\textbf{v}_{i}\cdot\textbf{E}))\frac{1}{\gamma_{i}^{2}}$,
but it gives no further usage.

Ones again, we consider the monopole approximation of the electric field
$E^{a}(\textbf{r}+\mbox{\boldmath $\xi$},t)\approx E^{a}(\textbf{r},t)$,
which basically gives the mean-field approximation,
and find
\begin{equation}\label{RHD2018Cl} \langle v_{i}^{a}E_{i}^{a}\rangle-\frac{1}{c^{2}}\langle v_{i}^{a}\textbf{v}_{i}^{2}E_{i}^{a}\rangle
=E^{a}(\textbf{r},t)\biggl(j^{a}-\frac{1}{c^{2}}Q^{a}\biggr), \end{equation}
where
\begin{equation}\label{RHD2018Cl Q a def} Q^{a}=\langle v_{i}^{a}\textbf{v}_{i}^{2}\rangle\end{equation}
is the flux vector of the velocity square which coincides (proportional) with the kinetic energy current in the nonrelativistic regime.

Finally, we present the evolution equation for the Gamma function
\begin{equation}\label{RHD2018Cl equation of Gamma evolution 2 monopole}
\partial_{t}\Gamma+\partial_{a}\Theta^{a}= -\frac{1}{c^{2}}\textbf{E}\biggl(\textbf{j}-\frac{1}{c^{2}}\textbf{Q}\biggr),
\end{equation}
where
we see that the Theta function is the current of the Gamma function
(as it is mentioned above judging on the structure of the definitions).
Hence, the set of equation partially closes itself.
So, evolution of Gamma function leads to functions $j^{a}$, $\Theta^{a}$,
which are introduced above, and single new function $Q^{a}$.

\subsubsection{Equation for the Theta function evolution}

Next equation which appears in the developing model is the evolution equation for the hydrodynamic Theta function.

Differentiating function (\ref{RHD2018Cl Theta function definition}) with respect to time we find the required equation.
Derivative of function (\ref{RHD2018Cl Theta function definition}) leads to three terms:
the derivative of the velocity,
the derivative of the delta function,
and the derivative of the reverse gamma factor.

The derivative of the delta function leads to the flux of the hydrodynamic Theta function
which is presented by function $\Xi^{ab}$ (\ref{RHD2018Cl Xi function definition}).

The derivative of the reverse gamma factor is considered at the derivation of the hydrodynamic Gamma function evolution:
$\partial_{t}(\gamma_{i}^{-1})$$=-\frac{e}{mc^{2}}\gamma_{i}^{-2}(\textbf{v}_{i}\cdot\textbf{E}_{i})$.

The acceleration provides the following term
$\gamma_{i}^{-1}\dot{v}_{i}^{a}=$
$(e/m)\gamma_{i}^{-2}[E_{i}^{a}+\varepsilon^{abc}v_{i}^{b}B_{i}^{c}/c-v_{i}^{a}v_{i}^{b}E_{i}^{b}/c^{2}]$.
The coefficient $\gamma_{i}^{-2}$ splits on two terms.
Hence, the term caused by the acceleration $\gamma_{i}^{-1}\dot{v}_{i}^{a}$ splits on six terms.

We present the evolution equation for the hydrodynamic Theta function
in the monopole (mean-field) approximation of the electromagnetic field
$$\partial_{t}\Theta^{a} +\partial_{b}\Xi^{ab}=\frac{e}{m}E^{a}\biggl[n-\frac{\Pi^{bb}}{c^{2}}\biggr]$$
\begin{equation}\label{RHD2018Cl equation of Theta evolution}
+\frac{e}{mc}\varepsilon^{abc}\biggl[j^{b}-\frac{Q^{b}}{c^{2}}\biggr]B^{c}-\frac{2e}{mc^{2}}E^{b}\biggl[\Pi^{ab}-\frac{L^{abcc}}{c^{2}}\biggr],\end{equation}
where
\begin{equation}\label{RHD2018Cl L fourth order definition} L^{abcd}=\langle v_{i}^{a}v_{i}^{b}v_{i}^{c}v_{i}^{d}\rangle. \end{equation}

It is essential to point out that
the evolution of $\Theta^{a}$ is mostly expressed via the concentration $n$, the current $j^{a}$, the flux of current $\Pi^{ab}$.
It means that the set equations is almost closing itself one again.
One additional function appears at this step $L^{abcd}$ (\ref{RHD2018Cl L fourth order definition})
along with function $Q^{a}$ (\ref{RHD2018Cl Q a def}) obtained at the derivation of equation
(\ref{RHD2018Cl equation of Gamma evolution 2 monopole}).
Therefore, we can try to stop at this step and truncate the set of equations.

\subsubsection{Xi function evolution}

The particle current evolution equation (\ref{RHD2018Cl current derivative Euler equation explicit})
suggests that we need evolution equation for the Xi function $\Xi^{ab}$.
However, it is necessary to create a limited set of equations.
Hence, at this stage of the model development find an equation of state for the Xi function $\Xi^{ab}$
(\ref{RHD2018Cl Xi function definition}).

\subsection{Equations of electromagnetic field}

The mean-field electromagnetic field $\textbf{E}$ and $\textbf{B}$
appearing in equations (\ref{RHD2018Cl current derivative Euler equation explicit}),
(\ref{RHD2018Cl equation of Gamma evolution 2 monopole}),
and (\ref{RHD2018Cl equation of Theta evolution})
satisfies the Maxwell equations
\begin{equation}\label{RHD2018Cl div B} \nabla \cdot\textbf{B}=0,\end{equation}
\begin{equation}\label{RHD2018Cl rot E}
\nabla\times \textbf{E}=-\frac{1}{c}\partial_{t}\textbf{B}, \end{equation}
\begin{equation}\label{RHD2018Cl div E} \nabla \cdot\textbf{E}=4\pi\sum_{s=e,i}e_{s}n_{s},\end{equation}
and
\begin{equation}\label{RHD2018Cl rot B with time}
\nabla\times \textbf{B}=\frac{1}{c}\partial_{t}\textbf{E}+\frac{4\pi}{c}\sum_{s=e,i}e_{s}\textbf{j}_{s}.\end{equation}

Full set of hydrodynamic equations is developed for each species,
while the superposition of all particles appears as the source of the electromagnetic field.
Hence, the sum of species "s" is presented in equations (\ref{RHD2018Cl div E}) and (\ref{RHD2018Cl rot B with time}).
as an example we show the summation on two species:
the electrons and ions.


Quantum analog of this concept is suggested in Ref. \cite{Maksimov QHM 99}.
The many-particle quantum hydrodynamics is developed for the number of physical systems
\cite{Andreev JPP 21}, \cite{Andreev Ch 21}, \cite{Andreev PoF 21}.

\section{Velocity field in equations for the relativistic hydrodynamics}

Introduce the velocity field in the traditional way as the ratio between the particle current and the concentration $\textbf{v}=\textbf{j}/n$.
Next, we need to recognize the contribution of the velocity field in other hydrodynamic functions.
The velocity field is the local average velocity or in other words it is the average velocity of all particle in the delta vicinity $\Delta$ of point $\textbf{r}$.
Therefore, we can split the velocity of each particle on the average velocity (the velocity field) $\textbf{v}$
and the deviation from the velocity field $\textbf{u}_{i}$ caused by the difference of velocities of particles related to the thermal effects: $\textbf{v}_{i}=\textbf{v}+\textbf{u}_{i}$.
Hence, function $\textbf{u}_{i}$ can be interpreted as the local thermal velocity of particles.
The definitions of the current (\ref{RHD2018Cl particle current definition}) and the velocity field show that the average of the thermal velocity is equal to zero $\langle\textbf{u}_{i}\rangle=0$.

We substitute the decomposition of the velocities in the definition of the hydrodynamic functions.
We start with the current flux
\begin{equation}\label{RHD2018Cl} \Pi^{ab}=\langle v_{i}^{a}v_{i}^{b}\rangle=nv^{a}v^{b}+p^{ab},\end{equation}
which  gives the flux of the current of particles on the thermal velocities $p^{ab}=\langle u_{i}^{a}u_{i}^{b}\rangle$
which is an analog of pressure,
but it is a different function.
The terms linear on the thermal velocity go to zero.
Next, we consider the hydrodynamic Theta function
\begin{equation}\label{RHD2018Cl}
\Theta^{a}=\biggl\langle \frac{v_{i}^{a}}{\gamma_{i}}\biggr\rangle=\Gamma v^{a}+t^{a},\end{equation}
where
$t^{a}=\langle\frac{u_{i}^{a}}{\gamma_{i}}\rangle$,
with
\begin{equation}\label{RHD2018Cl}
\gamma_{i}=\frac{1}{\sqrt{1-\frac{[\textbf{v}^{2}+2\textbf{v}\cdot\textbf{u}_{i}+\textbf{u}_{i}^{2}]}{c^{2}}}}.\end{equation}
Function $t^{a}$ can be called the thermal part of the hydrodynamic Theta function.
However, function $t^{a}$ also contains the velocity field in nonadditive form.
Same is true for the hydrodynamic Gamma function $\Gamma$
since "averaged" reverse gamma factor contains both the velocity field and the thermal velocity.

we also describe the structure of the hydrodynamic Xi function $\Xi^{ab}$ in terms of the velocity field:
\begin{equation}\label{RHD2018Cl}
\Xi^{ab}=\biggl\langle \frac{v_{i}^{a}v_{i}^{b}}{\gamma_{i}}\biggr\rangle
=\Gamma v^{a}v^{b}+ v^{a}t^{b} +t^{a}v^{b} +t^{ab},\end{equation}
where
$t^{ab}=\langle\frac{u_{i}^{a}u_{i}^{b}}{\gamma_{i}}\rangle$.

We ready to represent continuity equation and the particle current evolution equation,
but the hydrodynamic Gamma function evolution equation contains vector $Q^{a}$
and the hydrodynamic Theta function evolution equation includes vector $Q^{a}$ and the partial trace of tensor $L^{abcd}$.

here, we study the structure of presented functions:
$$Q^{a}=\langle v_{i}^{a}v_{i}^{b}v_{i}^{b}\rangle$$
\begin{equation}\label{RHD2018Cl}=nv^{a}v^{b}v^{b}+v^{a}p^{bb}+2v^{b}p^{ab}+q^{a},\end{equation}
where
\begin{equation}\label{RHD2018Cl}q^{a}=\langle u_{i}^{a}u_{i}^{b}u_{i}^{b}\rangle,\end{equation}
and
$$L^{abcd}=\langle v_{i}^{a}v_{i}^{b}v_{i}^{c}v_{i}^{d}\rangle
=nv^{a}v^{b}v^{c}v^{d}$$
\begin{equation}\label{RHD2018Cl}+[v,v,p]^{a,b,cd} +[v,q]^{a,bcd} +M^{abcd}.\end{equation}
We have rather huge expression for tensor $L^{abcd}$ in terms of the velocity field.
Therefore, we introduce the following notations including  permutations of similar terms.
\begin{equation}\label{RHD2018Cl}[v,q]^{a,bcd}\equiv v^{a}q^{bcd}+v^{b}q^{acd}+v^{c}q^{abd}+v^{d}q^{abc},\end{equation}
and
$$[v,v,p]^{a,b,cd}\equiv v^{a}v^{b}p^{cd}+v^{a}v^{c}p^{bd}$$
\begin{equation}\label{RHD2018Cl}+v^{a}v^{d}p^{bc} +v^{b}v^{c}p^{ad}+v^{b}v^{d}p^{ac}
+v^{c}v^{d}p^{ab}.\end{equation}
Above, we also introduce the third and fourth rank tensors composed of the thermal velocities:
\begin{equation}\label{RHD2018Cl}q^{abc}=\langle u_{i}^{a}u_{i}^{b}u_{i}^{c}\rangle,\end{equation}
and
\begin{equation}\label{RHD2018Cl}M^{abcd}=\langle u_{i}^{a}u_{i}^{b}u_{i}^{c}u_{i}^{d}\rangle.\end{equation}
Equation for the hydrodynamic Theta function evolution contains the partial trace of tensor $L^{abcc}$,
which has the following form:
$$L^{abcc}=nv^{a}v^{b}\textbf{v}^{2} +v^{a}v^{b}p^{cc} +2v^{a}v^{c}p^{bc}+2v^{b}v^{c}p^{ac}$$
\begin{equation}\label{RHD2018Cl}+v^{2}p^{ab} +v^{a}q^{b}+v^{b}q^{a}+2v^{c}q^{abc} +M^{abcc}.\end{equation}



\subsection{Intermediate form of the suggested hydrodynamic model}

The representation of the hydrodynamic functions via the velocity is given.
Therefore, we can represent the hydrodynamic equations (\ref{RHD2018Cl continuity equation}), (\ref{RHD2018Cl current derivative}),
(\ref{RHD2018Cl equation of Gamma evolution 2 monopole}), and (\ref{RHD2018Cl equation of Theta evolution})
via the velocity field.

The continuity equation has the traditional form:
\begin{equation}\label{RHD2018Cl continuity equation via velocity}
\partial_{t}n+\nabla\cdot (n\textbf{v})=0. \end{equation}

Equation of the velocity field evolution:
$$n(\partial_{t}+(\textbf{v}\cdot\nabla))v^{a}+\partial_{b}p^{ab}
=\frac{e}{m}\Gamma E^{a}+\frac{e}{mc}\Gamma\varepsilon^{abc}v^{b}B^{\gamma}$$
\begin{equation}\label{RHD2018Cl velocity field evolution equation}
+\frac{e}{mc}\varepsilon^{abc}t^{b}B^{\gamma}-\frac{e}{mc^{2}}(\Gamma v^{a}v^{b}+v^{a}t^{b}+t^{a}v^{b}+t^{ab})E^{b}
\end{equation}
has familiar left-hand side
while the right-hand side (at leats first three terms) has an intuitively understandable structure,
but new (untraditional) functions define the right-hand side.

Tensor $p^{ab}$ looks like the pressure.
However, the traditional pressure is the momentum transmitted through the element of surface during the unit of time.
While $p^{ab}$ is the current of particles transmitted through the element of surface during the unit of time.

Next, we find equation for the hydrodynamic Gamma function
$$\partial_{t}\Gamma +\partial_{b}(v^{b}\Gamma)+\partial_{b}t^{b} =-\frac{e}{mc^{2}}\biggl[nv^{b}E_{b}$$
\begin{equation}\label{RHD2018Cl evolution of Gamma via velocity}
-\frac{1}{c^{2}}E^{b}(nv_{b}\textbf{v}^{2} +v_{b}p^{cc}+2p^{bc}v_{c}+q_{b})\biggr]. \end{equation}
Finally, we present the hydrodynamic Theta function evolution equation
$$\partial_{t}t^{a} +\partial_{b}(t^{a}v^{b})+t^{b}\partial_{b}v^{a}+\partial_{b}t^{ab}$$
$$-\frac{e}{mc^{2}}v^{a}\biggl[nv^{b}E_{b}
-\frac{1}{c^{2}}E^{b}(nv_{b}\textbf{v}^{2} +v_{b}p^{cc}+2p^{bc}v_{c}+q_{b})\biggr]$$
$$-\frac{\Gamma}{n}\partial_{b}p^{ab}
+\frac{e}{m}\frac{\Gamma}{n}\biggl[\Gamma E^{a} +\frac{1}{c}\Gamma\varepsilon^{abc}v_{b}B_{c}-\frac{1}{c^{2}}\Gamma v^{a}v^{b}E_{b}$$
$$+\frac{1}{c}\varepsilon^{abc}t_{b}B_{c} -\frac{1}{c^{2}}(v^{a}t^{b}+t^{a}v^{b}+t^{ab})E_{b}\biggr]
=\frac{e}{m}nE^{a}$$
$$+\frac{e}{mc}n\varepsilon^{abc}v_{b}B_{c} -\frac{2e}{mc^{2}}(nv^{a}v^{b}+p^{ab})E_{b}
-\frac{e}{mc^{2}}E^{a}(n\textbf{v}^{2}+p^{bb})$$
$$-\frac{e}{mc^{2}}\frac{1}{c}\varepsilon^{abc}(nv_{b}\textbf{v}^{2}+v_{b}p^{dd}+2v^{d}p_{bd}+q_{b})B_{c}$$
$$+\frac{2e}{mc^{2}}\frac{1}{c^{2}}[nv^{a}v^{b}\textbf{v}^{2}+v^{a}v^{b}p^{cc}+p^{ab}\textbf{v}^{2}+2p^{ac}v^{b}v_{c}+2v^{a}p^{bc}v_{c}$$
\begin{equation}\label{RHD2018Cl evolution of Theta via velocity}
+v^{a}q^{b}+v^{b}q^{a}+2v^{c}q^{abc}+M^{abcc}]E_{b}. \end{equation}

Equations (\ref{RHD2018Cl continuity equation via velocity})-(\ref{RHD2018Cl evolution of Gamma via velocity})
are the basic set of relativistic hydrodynamic equations of the ideal plasmas suggested in this paper.
However, equations (\ref{RHD2018Cl continuity equation via velocity})-(\ref{RHD2018Cl evolution of Gamma via velocity})
should be truncated to obtain a closed set of equations applicable for the study of plasmas.

This paper is focused on the method of derivation of hydrodynamic equations for the relativistic plasmas.
However, these equations have been presented in Refs.
\cite{Andreev 2021 05}, \cite{Andreev 2021 06}, \cite{Andreev 2021 07}, \cite{Andreev 2021 08},
including waves in the magnetized plasmas \cite{Andreev 2021 06}, \cite{Andreev 2021 07}, \cite{Andreev 2021 08}.

\section{Truncation}

The truncation requires equations of state for several functions including
$\Pi^{ab}$ (reducing to $p^{ab}$), $\Xi^{ab}$ (reducing to $t^{ab}$),
$Q^{a}$ (reducing to $q^{a}$), $L^{abcd}$ (reducing to $M^{abcd}$).
They should be presented via functions $n$, $v^{a}$, $\Gamma$, $t^{a}$.

Explicit microscopic definitions like (\ref{RHD2018Cl concentration definition}), (\ref{RHD2018Cl Gamma function definition}),
(\ref{RHD2018Cl Theta function definition}), (\ref{RHD2018Cl Xi function definition}) are not useful for calculation of equilibrium values of considering functions or derivations of equations of state.
These definitions are based on the exact microscopic motion of particles
which is unknown.
However, the approximate application of the equations for the higher rank tensors can give this information.
However, we have not developed extended model,
it is planed as the future work.
Here we go another way and neglect the physical picture demonstrated during derivation.
So, some formal technic is applied to get the required equations of state.
This situation is usual for the hydrodynamic models of classical fluids and plasmas.
Required values can be obtained via the equilibrium distribution functions $f_{0}(p)$,
where $p$ is the module of momentum.

For instance the Gamma function is the average of the reverse of the relativistic factor $\gamma_{i}$.
Hence, the equilibrium Gamma unction can be calculated as
$\Gamma_{0}=\int\gamma^{-1}(p)f_{0}(p)d^{3}p$,
where
$\gamma(p)=\sqrt{p^{2}+m^{2}c^{2}}/mc$.
Here we make substitution of concepts.
We use the distribution function instead of the arithmetic average on the $\Delta$-vicinity.

\subsection{Equations of state: rest frame}

Calculate required expressions using relativistic equilibrium distribution function presented in the rest frame
\begin{equation}\label{RHD2018Cl eq rest frame distribution}
f_{0r}(p)=\textrm{Z} e^{-\epsilon/T},
\end{equation}
where
the subindex $0$ refers to the equilibrium state,
the subindex $r$ refers to the rest frame,
\begin{equation}\label{RHD2018Cl}
\textrm{Z}=\frac{n}{4\pi m^{2}cTK_{2}(\frac{mc^{2}}{T})},
\end{equation}
with $T$ is the equilibrium temperature in the energy units,
$p$ is the momentum,
$K_{2}(b)$ is the second order Macdonald function,
and $\epsilon=\sqrt{m^{2}c^{4}+p^{2}c^{2}}$.
The rest frame is the inertial frame,
where the fluid is macroscopically motionless or it has minimal energy.
The energy momentum tensor presented in the Introductions is constructed in the first regime for the macroscopically motionless fluid.
Same assumption is made for the distribution function (\ref{RHD2018Cl eq rest frame distribution}).
The useful application of the rest frame includes the assumption
that
the deviations from the macroscopically motionless state are relatively small.
So, we can consider the medium as the average background,
where all structures and flows can be viewed as processes happening on this background.
Hence, we introduce some equilibrium-like background for the large amplitude nonlinear processes.
The rest frame can be easily associated with this background.
If we have infinite number of the large amplitude structures overlapping each over we have no distinguishable background.
In this case the rest frame can be found as the inertial frame,
where system have minimal energy.
However, this formal step does not give simple physical picture to include in calculations of the distribution function or the energy-momentum tensor.
The Macdonald functions $K_{\mu}(b)$ have the following definition (the integral form):
\begin{equation}\label{RHD2018Cl}
K_{\mu}(b)=\frac{\sqrt{\pi}b^{\mu}}{\Gamma(\mu+\frac{1}{2})}\int_{1}^{+\infty}(t^{2}-1)^{\mu-1/2}e^{-bt}dt,
\end{equation}
where $Re\mu>-1/2$ and $Re b>0$.

For instance, the Macdonald functions $K_{\mu}(b)$ allow to calculate analytically the equilibrium value of the average reverse $\gamma$-factor:
$\Gamma_{0}=m^{2}cT\textrm{Z}K_{1}(b)$,
where $b=mc^{2}/T$.

Presented analysis provides the following equations of state
$p^{ab}=U_{p}^{2}n\delta^{ab}$,
$t^{ab}=U_{t}^{2}n\delta^{ab}$,
$q^{a}=0$,
$M^{abcd}=(U_{M}^{4}/3) nI_{0}^{abcd}$,
where $I_{0}^{abcd}=\delta^{ab}\delta^{cd}+\delta^{ac}\delta^{bd}+\delta^{ad}\delta^{bc}$,
and $U_{p}$, $U_{t}$, $U_{M}$ are constants.

After the application of the equations of state for the high-rank tensors introduced in the model
we obtain the following set of truncated hydrodynamic equations.
The continuity equation:
\begin{equation}\label{RHD2018Cl continuity equation via velocity Truncated}
\partial_{t}n+\nabla\cdot (n\textbf{v})=0. \end{equation}
Equation of the velocity field evolution:
$$n(\partial_{t}+(\textbf{v}\cdot\nabla))\textbf{v}+U_{p}^{2}\nabla n
=\frac{e}{m}\Gamma \textbf{E}$$
$$+\frac{e}{mc}\Gamma [\textbf{v}\times \textbf{B}]
+\frac{e}{mc} [\textbf{t}\times \textbf{B}]
-\frac{e}{mc^{2}}\biggl(
\Gamma \textbf{v} (\textbf{v}\cdot\textbf{E})$$
\begin{equation}\label{RHD2018Cl velocity field evolution equation Truncated}
+\textbf{v}(\textbf{t}\cdot\textbf{E})
+\textbf{t}(\textbf{v}\cdot\textbf{E})
+U_{t}^{2} n \textbf{E}\biggr).
\end{equation}
The Gamma function evolution equation
$$\partial_{t}\Gamma +\nabla\cdot(\textbf{v}\Gamma)+\nabla\cdot \textbf{t} =-\frac{e}{mc^{2}}\biggl[n\textbf{v}\cdot\textbf{E}$$
\begin{equation}\label{RHD2018Cl evolution of Gamma via velocity Truncated}
-\frac{1}{c^{2}}(n(\textbf{v}\cdot\textbf{E}) \textbf{v}^{2} +5U_{p}^{2}n \textbf{v}^{2})\biggr]. \end{equation}
The Theta function evolution equation
$$\partial_{t}\textbf{t} +(\textbf{v}\cdot\nabla)\textbf{t}+\textbf{t}(\nabla\cdot\textbf{v})+(\textbf{t}\cdot\nabla)\textbf{v}+U_{t}^{2}\nabla n$$
$$-\frac{e}{mc^{2}}v^{a}\biggl[n\textbf{v}\cdot \textbf{E}
-\frac{1}{c^{2}}\biggl(n(\textbf{v}\cdot\textbf{E}) \textbf{v}^{2} +(\textbf{v}\cdot\textbf{E}) 5U_{p}^{2}n\biggr)\biggr]$$
$$-\frac{\Gamma}{n}U_{p}^{2}\nabla n
+\frac{e}{m}\frac{\Gamma}{n}\biggl[\Gamma \textbf{E} +\frac{1}{c}\Gamma [\textbf{v}\times\textbf{B}]
-\frac{1}{c^{2}}\Gamma \textbf{v}(\textbf{v}\cdot\textbf{E})$$
$$+\frac{1}{c} [\textbf{t}\times \textbf{B}]
-\frac{1}{c^{2}}(\textbf{v}(\textbf{t}\times\textbf{E})+\textbf{t}(\textbf{v}\times\textbf{E})+U_{t}^{2}n\textbf{E})\biggr]
$$
$$=\frac{e}{m}n\textbf{E}
+\frac{e}{mc}n[\textbf{v}\times \textbf{B}]
-\frac{2e}{mc^{2}}(n\textbf{v}(\textbf{v}\times\textbf{E})+U_{p}^{2}n\textbf{E})$$
$$-\frac{e}{mc^{2}}\textbf{E}(n\textbf{v}^{2}+3U_{p}^{2}n)
-\frac{e}{mc^{2}}\frac{1}{c}(n\textbf{v}^{2}+5U_{p}^{2}n)[\textbf{v}\times\textbf{B}]$$
$$+\frac{2e}{mc^{2}}\frac{1}{c^{2}}
[n\textbf{v}(\textbf{v}\cdot\textbf{E}) \textbf{v}^{2}
+5U_{p}^{2}n\textbf{v}(\textbf{v}\cdot\textbf{E})
$$
\begin{equation}\label{RHD2018Cl evolution of Theta via velocity Truncated}
+U_{p}^{2}n\textbf{E} \textbf{v}^{2}
+2\textbf{v} U_{p}^{2}n(\textbf{v}\cdot \textbf{E})
+\frac{5}{3}U_{M}^{4}\textbf{E}]. \end{equation}

Equations (\ref{RHD2018Cl continuity equation via velocity Truncated})
-(\ref{RHD2018Cl evolution of Theta via velocity Truncated})
are coupled to the Maxwell equations (\ref{RHD2018Cl div B})-(\ref{RHD2018Cl rot B with time}).

Three different characteristic velocities $U_{p}$, $U_{t}$, and $U_{M}$ are calculated in the following way.

The application of the isotropic distribution function leads to diagonal form of tensors $p^{ab}$ and $t^{ab}$:
$p^{ab}=p\delta^{ab}$ and $t^{ab}=\tilde{t}\delta^{ab}$.
The "diagonal" form is obtained for tensor $M^{abcd}$ as well:
$M^{abcd}=M_{0}(\delta^{ab}\delta^{cd}+\delta^{ac}\delta^{bd}+\delta^{ad}\delta^{bc})/3$.
The equilibrium expressions for functions
$p$, $\tilde{t}$, $\textbf{q}$, $M$ are used as the equations of state for the nonequilibrium functions.
Approximate calculation of functions $p^{ab}$, $t^{ab}$, $\textbf{q}$, and $M^{abcd}$ gives to the following representations
$p^{ab}=c^{2}\delta^{ab}\tilde{Z}f_{1}(\beta)/3$,
$t^{ab}=c^{2}\delta^{ab}\tilde{Z}f_{2}(\beta)/3$
$M^{abcd}=c^{4}(\delta^{ab}\delta^{cd}+\delta^{ac}\delta^{bd}+\delta^{ad}\delta^{bc})\tilde{Z}f_{3}(\beta)/15$,
and
$\textbf{q}=0$,
where
$\beta=mc^{2}/T$,
$\tilde{Z}=4\pi Z (mc)^{3}=n\beta K_{2}^{-1}(\beta)$,
\begin{equation}\label{RHD2021ClLM f 1} f_{1}(\beta)=\int_{1}^{+\infty}\frac{d x}{x}(x^{2}-1)^{3/2}e^{-\beta x}, \end{equation}
\begin{equation}\label{RHD2021ClLM f 2} f_{2}(\beta)=\int_{1}^{+\infty}\frac{d x}{x^{2}}(x^{2}-1)^{3/2}e^{-\beta x}, \end{equation}
and
\begin{equation}\label{RHD2021ClLM f 3} f_{3}(\beta)=\int_{1}^{+\infty}\frac{d x}{x^{3}}(x^{2}-1)^{5/2}e^{-\beta x}. \end{equation}
Functions $f_{1}(\beta)$, $f_{2}(\beta)$ and $f_{3}(\beta)$ are calculated numerically below for the chosen values of temperatures.
For each function describing the thermal we introduce corresponding velocity
$\delta p=U_{p}^{2} \delta n$,
$\delta \tilde{t}=U_{t}^{2} \delta n$,
$\delta M=U_{M}^{4} \delta n$.

\subsection{Equations of state: arbitrary inertial frame}

For the calculation of equations of state
we have used the relativistic Maxwellian distribution function in the rest frame.
This approach narrows down the variety of physical scenario,
where the suggested model can be applied.
Let us describe some backgrounds for the further generalizations.
First, we can use the relativistic Maxwellian distribution in the arbitrary frame:
\begin{equation}\label{RHD2018Cl distribution p A}
f_{0}(p)=\hat{Z} e^{U_{\alpha}P^{\alpha}/T},
\end{equation}
where
$P^{\alpha}=p^{\alpha}+eA^{\alpha}$ is the canonical momentum,
$p^{\alpha}=\{\varepsilon/c,\textbf{p}\}$ is the four-momentum,
$A^{\alpha}=\{\phi,\textbf{A}\}$ is the four-potential of the electromagnetic field,
$U_{\alpha}=\{-c,\textbf{v}\}$
(while $U^{\alpha}=\{c,\textbf{v}\}$)
is the hydrodynamic four-velocity field giving the four-current $j^{\alpha}=nU^{\alpha}$,
$\textbf{v}$ is the hydrodynamic velocity field,
and
\begin{equation}\label{RHD2018Cl Z hat}
\hat{Z}=\frac{n}{4\pi m^{2}cTK_{2}(\frac{mc^{2}}{T})} e^{e\phi/T}.
\end{equation}
with $A^{0}=\Phi$
The hydrodynamic four-velocity field entering the distribution function along with the concentration presented in $\hat{Z}$.

More general distribution functions are used in literature for the derivation of hydrodynamics from the kinetic model.
For instance, generalized prefactor in front of exponential function is discussed in Ref. \cite{Hazeltine APJ 2002},
where the prefactor is constructed of several hydrodynamic functions in the covariant form.
This function is aimed to cover the gyrokinetic effects in the relativistic plasmas.

\section{On the covariant structure of hydrodynamic equations}

At the phenomenological derivations of the macroscopic equations of motion
we need to use some fundamental principles
which allows to avoid some misinterpretations.
One of such fundamental principles in relativistic physics is the covariant form of equations.
However, sometimes one needs to neglect the covariant form of equations in order to find an appropriate macroscopic in a particular frame.
One of the most important examples is the quantum field theory at the finite temperatures,
where there is chosen reference frame bound to the thermostat.

In our derivation
we are not bound to particular inertial frame.
we start our derivation in the arbitrary inertial frame.
However, afterwords we work in this microscopically fixed frame in order to make transition to the macroscopic scale.
Systematic application of the equations correctly describing the microscopic motion allows us to obtain the correct macroscopic equations.

We use microscopic equation of motion presented in noncovariant form (\ref{RHD2018Cl Newton eq}).
Consequently, the obtained macroscopic equations appear in noncovariant notations.
Moreover, presented equations are found in the three-vector notations.
Nevertheless, we can use these equations for the analysis of relativistic effects working in the single inertial frame.

Introduced above hydrodynamic functions are presented in the three-vector notations.
However, they can be combined as components of the four-tensors:
\begin{equation}\label{RHD2018Cl four tensor Pi} \Pi^{\alpha\beta}=\left(
                                                      \begin{array}{cc}
                                                        n c^{2} & n v^{b} c \\
                                                        nv^{a} c & \Pi^{ab} \\
                                                      \end{array}
                                                    \right),
\end{equation}
and
\begin{equation}\label{RHD2018Cl four tensor Gamma} \Gamma^{\alpha\beta}=\left(
                                                      \begin{array}{cc}
                                                        \Gamma c^{2} & t^{b} c \\
                                                        t^{a} c & t^{ab} \\
                                                      \end{array}
                                                    \right).
\end{equation}
These are analogs of the energy-momentum tensor
which is describe below at the analysis of the four-momentum density evolution.

\section{On a possible structure of the energy-momentum tensor}

Presented method of the microscopic derivation allows to derive the energy-momentum four-vector
and derive equation for its evolution providing the energy-momentum tensor.
This equation is not a part of suggested hydrodynamic equations
because the found structure of hydrodynamic equations is a consequence revealed by the concentration evolution.

Relativistic hydrodynamic allows several generalization of concentration and velocity field.
Such as the particle current and the momentum density coincide in nonrelativistic limit,
but they are differen functions in general case.

Equations (\ref{RHD2018Cl continuity equation via velocity})-(\ref{RHD2018Cl evolution of Gamma via velocity}) show
that the energy and momentum do not appear as a part of evolution of concentration and current.
While the concentration and current are relevant since
they exist as sources of field in the Maxwell equations.

However, we show the energy-momentum $\{\epsilon/c, p^{a}\}$ evolution
to demonstrate agreement between presented method and relativistic hydrodynamics presented in other works.
Let us to point out  the general matrix structure of the energy-momentum tensor together with its microscopic definitions:
\begin{equation}\label{RHD2018Cl T a b matrix structure} T^{\alpha\beta}=\left(
                                                      \begin{array}{cc}
                                                        \epsilon & p^{b}c \\
                                                        p^{a}c & T^{ab} \\
                                                      \end{array}
                                                    \right)
=\langle v_{i}^{\alpha}p_{i}^{\beta}\rangle.
 \end{equation}
to the best of my knowledge all relativistic hydrodynamic models are based on the momentum balance equation.
Therefore, we present the microscopic definition of the momentum density
\begin{equation}\label{RHD2018Cl momentum density definition}
p^{\alpha}(\textbf{r},t)=\frac{1}{\Delta}\int_{\Delta}d\mbox{\boldmath $\xi$}\sum_{i=1}^{N} p^{\alpha}_{i}(t)
\delta(\textbf{r}+\mbox{\boldmath $\xi$}-\textbf{r}_{i}(t)),
\end{equation}
where $p^{\alpha}_{i}(t)=\{\epsilon_{i}/c,\textbf{p}_{i}\}=\{mc\gamma_{i},m\gamma_{i}\textbf{v}_{i}\}$.

Next, we consider the time evolution of the momentum density via the calculation of the time derivative of the presented function
$$\partial_{t}p^{a}=\frac{1}{\Delta}\int_{\Delta}d\mbox{\boldmath $\xi$}\sum_{i=1}^{N}
\biggl(q_{i}E_{i}^{a}+\frac{q_{i}}{c}\varepsilon^{abc}v_{i}^{b}B_{i}^{c}\delta_{i}\biggr)$$
\begin{equation}\label{RHD2018Cl momentum evolution time derivative}
-\partial_{b}\frac{1}{\Delta}\int_{\Delta}d\mbox{\boldmath $\xi$}\sum_{i=1}^{N}p_{i}^{a}v_{i}^{b}\delta_{i},\end{equation}
where $\delta_{i}=\delta(\textbf{r}+\mbox{\boldmath $\xi$}-\textbf{r}_{i}(t))$.

In the meanfield approximation equation (\ref{RHD2018Cl momentum evolution time derivative}) reduces to the following equation:
\begin{equation}\label{RHD2018Cl momentum evolution meanfield}
\partial_{t}p^{a}+\partial_{b}\langle mv_{i}^{a}v_{i}^{b}\gamma_{i}\rangle
=qnE^{a}+\frac{q}{c}n\varepsilon^{abc}v^{b}B^{c}. \end{equation}
the left-hand side of equation (\ref{RHD2018Cl momentum evolution meanfield}) can be rewritten via the energy-momentum tensor in the four-vector.
Therefore, the right-hand side should be rewritten in some notations via the tensor of electromagnetic field.
Equation (\ref{RHD2018Cl momentum evolution meanfield}) requires two equations of state.
First, one needs to get rid of the momentum density $p^{a}$ presenting it in terms of the velocity field
to make it in agreement with the right-hand side of momentum evolution equation (\ref{RHD2018Cl momentum evolution meanfield})
(which is represented via velocity),
the Maxwell equations and the continuity equation.

Moreover, the momentum flux $\langle v_{i}^{a}v_{i}^{b}\gamma_{i}\rangle$ (containing the pressure tensor) requires
an equation of state as the function of concentration and the velocity field.
Any hydrodynamic model requires one or several equations of state.

Our model shows that in the relativistic case there are three generalizations of the Euler equation.
they are equations for particle current (velocity field) $\textbf{j}\sim \textbf{v}$, t-vector function $\textbf{t}$, and the momentum $\textbf{p}$.
All of them have same nonrelativistic limit.

The microscopic energy momentum tensor (\ref{RHD2018Cl T a b matrix structure}) is derived
as a part of the momentum balance equation (\ref{RHD2018Cl momentum evolution meanfield}).
However, we show above that
the relativistic hydrodynamic can be contracted of several scalars $n$, $\Gamma$, $\tilde{t}$, $p$, $M_{0}$
and two vector fields $\textbf{v}$ and $\textbf{t}$.
It gives the following four-vector notations $u^{\alpha}=\{c,\textbf{v}\}$ and $\Gamma^{\alpha}=\{\Gamma, \textbf{t}\}$.
Hence, an attempt of the microscopic contraction of the energy-momentum tensor $T^{\alpha\beta}$ discussed in the Introduction
can be generalized up to account of term proportional to $\Gamma^{\alpha}\Gamma^{\beta}$
along with $u^{\alpha}u^{\beta}$ and $\delta^{\alpha\beta}$.
However, our further discussion demonstrate no necessity to extend the energy momentum tensor,
but the introduction of novel tensors $\Pi^{\alpha\beta}$ (\ref{RHD2018Cl four tensor Pi})
and $\Gamma^{\alpha\beta}$ (\ref{RHD2018Cl four tensor Gamma}).

\section{Longitudinal one-dimensional waves in the relativistic isotropic plasmas}

Let us demonstrate the application of the developed relativistic hydrodynamic model
(\ref{RHD2018Cl continuity equation via velocity Truncated})-(\ref{RHD2018Cl evolution of Theta via velocity Truncated})
on the relatively simple (but fundamentally important) examples.
We start our illustration with the high-frequency longitudinal waves in the isotropic plasmas.
We focus on the regime,
where plasma is assumed macroscopically motionless,
but this plasmas has the relativistic temperature $T\sim m_{e}c^{2}$.
The temperature is proportional to the trace of pressure,
which is a part of the momentum flux tensor $T^{ab}=\langle v_{i}^{a}p_{i}^{b}\rangle$.
hence, the temperature can be larger then $m_{e}c^{2}$ due to the relativistic $\gamma$-factor i the definition of $T^{ab}$.

Considering system is described by nonzero equilibrium concentration $n_{0}$ and the equilibrium average $\gamma$-factor $\Gamma_{0}=n_{0}K_{1}(b)/K_{2}(b)$.
The equilibrium velocity field $\textbf{v}_{0}$, vector $\textbf{t}_{0}$, and the electric field $\textbf{E}_{0}$ are equal to zero.
The equilibrium magnetic field is equal to zero and perturbations of the magnetic field are absent.

We present the corresponding linearized equations set
which follows from equations
(\ref{RHD2018Cl continuity equation via velocity Truncated})-(\ref{RHD2018Cl evolution of Theta via velocity Truncated})
for the linear approximation on the small amplitude perturbations:
\begin{equation}\label{RHD2018Cl continuity equation lin 1D}
\partial_{t}\delta n+n_{0}\partial_{x} \delta v_{x}=0, \end{equation}
\begin{equation}\label{RHD2018Cl velocity field evolution equation lin 1D}
n_{0}\partial_{t}\delta v_{x}+\partial_{x}\delta p
=\frac{e}{m}\Gamma_{0} \delta E_{x}-\frac{e}{mc^{2}}\tilde{t}_{0}\delta E^{x},
\end{equation}
\begin{equation}\label{RHD2018Cl evolution of Gamma lin 1D}
\partial_{t}\delta\Gamma +\Gamma_{0}\partial_{x}\delta v_{x}+\partial_{x}\delta t_{x} =\frac{e}{mc^{2}}\frac{q_{0}^{x}}{c^{2}}\delta E^{x}, \end{equation}
and
$$\partial_{t}\delta t_{x} +\partial_{x}\delta \tilde{t}-\frac{\Gamma_{0}}{n_{0}}\partial_{x}\delta p +\frac{e}{m}\frac{\Gamma_{0}^{2}}{n_{0}}\delta E_{x}$$
\begin{equation}\label{RHD2018Cl evolution of Theta lin 1D}
=\frac{e}{m}n_{0}\delta E_{x} -5\frac{e}{mc^{2}}p_{0}\delta E_{x} +2\frac{e}{mc^{2}}M_{0}^{xxcc}\delta E_{x}. \end{equation}

Solve equations (\ref{RHD2018Cl continuity equation lin 1D})
and (\ref{RHD2018Cl velocity field evolution equation lin 1D})
to get $\delta n$ and $\delta v_{x}$ as functions of the electric field perturbations $\delta E_{x}$
and find the following expressions:
\begin{equation}\label{RHD2018Cl v x via E}
\delta v_{x}=\imath\omega\frac{e}{m}\frac{\Gamma_{0}-\frac{1}{c^{2}}\tilde{t}_{0}}{n_{0}(\omega^{2}-U_{p}^{2}k^{2})}\delta E_{x}, \end{equation}
\begin{equation}\label{RHD2018Cl concentration lin via E}
\delta n=\imath k\frac{e}{m}\frac{\Gamma_{0}-\frac{1}{c^{2}}\tilde{t}_{0}}{\omega^{2}-U_{p}^{2}k^{2}}\delta E_{x}. \end{equation}

Functions $\delta \Gamma$ and $\delta t_{x}$ can be also found as functions of the electric field perturbations $\delta E_{x}$ from equations
(\ref{RHD2018Cl evolution of Gamma lin 1D}) and (\ref{RHD2018Cl evolution of Theta lin 1D}),
but
these functions go nowhere.
They do not affect the evolution of the concentration $\delta n$ and the velocity field $\delta v_{x}$ in the linear approximation.
They do not contribute in the equations of field either.
The nonlinear evolution should be affected by these functions.
This conclusion is correct for the isotropic medium.
More complex picture would appear at the presence of the flows and/or the magnetic field.

We substitute the concentration (\ref{RHD2018Cl concentration lin via E}) in the Poisson equation $\nabla\cdot\textbf{E}=4\pi\rho$,
where $\rho$ is the charge density of the plasmas.
In our case it goes to
$\partial_{x}E_{x}=4\pi e\delta n$.
It gives the spectrum of the relativistic Langmuir waves:
\begin{equation}\label{RHD2018Cl Langm wave spectrum}
\omega^{2}=\omega_{Le}^{2}\frac{\Gamma_{0}-\frac{1}{c^{2}}\tilde{t}_{0}}{n_{0}} +U_{p}^{2}k^{2}.
\end{equation}

The Langmuir wave spectrum (\ref{RHD2018Cl Langm wave spectrum}) requires three equations of state,
while whole set of linearized equations includes more unknown functions approximately calculated for the truncation.

Necessary equations of state are demonstrated above,
where it is shown
that function $\tilde{t}_{0}$ is represented via the characteristic velocity $U_{t}$.
However, we want to discuss briefly its more general structure for different values of the velocity field $\textbf{v}$ contributing in $\tilde{t}_{0}$ in the general case.

Function $\tilde{t}$ has the following structure
in the arbitrary regime, where
the thermal velocity $\textbf{u}_{i}$ and the velocity field $\textbf{v}$ are relativistic and comparable to each other
\begin{equation}\label{RHD2018Cl tilde t arbitrary v and u}\tilde{t}=\biggl\langle\frac{u_{i}^{x}u_{i}^{x}}{\gamma_{i}}\biggr\rangle
=\biggl\langle u_{i}^{x}u_{i}^{x}\sqrt{1-\frac{\textbf{v}^{2}+2\textbf{u}_{i}\cdot\textbf{v}+\textbf{u}_{i}^{2}}{c^{2}}}\biggr\rangle.\end{equation}
Function $\tilde{t}$ reduces to the flux of particle current
If we dial with the relativistic beam (beam of electrons for instance),
where the thermal effects are neglegible
\begin{equation}\label{RHD2018Cl t via p cold}\tilde{t}=\sqrt{1-\frac{\textbf{v}^{2}}{c^{2}}}\cdot p,\end{equation}
where
the velocity field of the beam $\textbf{v}$ is also incorporated.

However, the opposite limit is more interesting.
If the thermal effects dominate over the velocity field existing in the relativistically hot beamless plasmas
function $\tilde{t}$ combines of fluxless part of $\tilde{t}$ and corrections caused by the velocity field:
\begin{equation}\label{RHD2018Cl expansion of t}\tilde{t}\approx\biggl\langle u_{i}^{x}u_{i}^{x} \sqrt{1-\frac{\textbf{u}_{i}^{2}}{c^{2}}}\biggr\rangle
-\frac{1}{c^{2}}v^{a}\biggl\langle u_{i}^{x}u_{i}^{x}u_{i}^{a} \frac{1}{\sqrt{1-\frac{\textbf{u}_{i}^{2}}{c^{2}}}}\biggr\rangle.\end{equation}



For the further analysis of spectrum (\ref{RHD2018Cl Langm wave spectrum})
we consider the main term in the expansion (\ref{RHD2018Cl expansion of t}).
Hence, we completely neglect the contribution of the velocity field in this function.

Finally, we have the following spectrum for the longitudinal waves in the macroscopically motionless isotropic plasmas:
\begin{equation}\label{RHD2018Cl Langmuir wave spectrum}
\omega^{2}=\omega_{Le}^{2}\biggl(\frac{K_{1}}{K_{2}}-\frac{U_{t}^{2}}{c^{2}}\biggr) +U_{p}^2 k^{2}.
\end{equation}
Fraction $\frac{K_{1}(b)}{K_{2}(b)}$ goes to 1 at the small temperatures (so the argument $b=mc^{2}/T$ goes to infinity),
since the argument of the Macdonald functions goes to infinity.

Numerical analysis of spectrum (\ref{RHD2018Cl Langmuir wave spectrum}) can be found in Ref. \cite{Andreev 2021 05}.
Nevertheless, we want to point out that additional multiplayer in front of the square of the Langmuir frequency is below unit.
Hence, the relativistic temperature effects reduces the minimal frequency of the minimal frequency of the longitudinal wave below $\omega_{Le}$.
This decrease can be of several orders.


\section{Relativistic beams in the nonrelativistic plasmas}

Major goal for the development of the presented hydrodynamic model is the study of plasmas with the large temperatures of order of $m_{e}c^{2}$ and above.
However, for justification of the model,
it is useful to consider relativistic effects in absence of the temperature effects.
This regime corresponds to the propagation of the relativistic beam through the plasmas.
Let us consider the high-frequency perturbations in the macroscopically motionless plasmas through which the electron beam propagates.
The beam is characterized by the equilibrium concentration $n_{0b}\ll n_{0e}$.
The equilibrium velocity of beam is $\textbf{v}_{0}=\{v_{0},0,0\}$.
We consider propagation of excitations parallel to the direction of the beam propagation $\textbf{k}=\{k,0,0\}$.
Under described conditions
the set of hydrodynamic equations
(\ref{RHD2018Cl continuity equation via velocity Truncated})-(\ref{RHD2018Cl evolution of Theta via velocity Truncated})
used for each species gives the following dispersion equation:
\begin{equation}\label{RHD2018Cl dispersion eq with beam}
1-\frac{\omega_{Le}^{2}\biggl(\frac{K_{1}}{K_{2}}-\frac{U_{t}^{2}}{c^{2}}\biggr)}{\omega^{2}-U_{p}^2 k^{2}} -\frac{\omega_{Lb}^{2}}{\gamma_{0}^{3}(\omega-kv_{0})^{2}}=0, \end{equation}
where $\gamma_{0}=\sqrt{1-v_{0}^{2}/c^{2}}^{-1}$,
and
$\omega_{Lb}^{2}=4\pi e^{2}n_{0b}/m_{e}$.
Presented cold regime corresponds to equation found in Refs.
\cite{Bret PoP 2006},
\cite{Bret PPCF 06},
\cite{Bret ApJ 09},
\cite{Bret PoP 2016},
\cite{Akhiezer book 74}.

\section{Plane waves in the relativistically hot magnetized plasmas}

\subsection{Linearly polarized waves propagating parallel to the magnetic field}

Similarly to the Sec. VII
we consider the plasmas
which are macroscopically motionless in the equilibrium state.
However, here we consider the plasmas placed in the uniform constant magnetic field.
The equilibrium state is characterized by the constant nonzero values for the concentration $n_{0}$,
the equilibrium average $\gamma$-factor $\Gamma_{0}$,
and the magnetic field $\textbf{B}_{0}=B_{0}\textbf{e}_{z}$.

We consider waves propagating parallel to the external magnetic field $\textbf{k}=\{0,0,k\}\parallel \textbf{B}_{0}$
with the electric field perturbations parallel to the external magnetic field $\delta \textbf{E}=\{0,0,E_{z}\}\parallel \textbf{B}_{0}$ as well.
We focus on the small amplitude excitations and consider they linear dynamic.
In this regime the set of hydrodynamic equations
(\ref{RHD2018Cl continuity equation via velocity Truncated})-(\ref{RHD2018Cl evolution of Theta via velocity Truncated})
reduces to the following couple of equations:
\begin{equation}\label{RHD2018Cl eq v z lin with B} -\imath\omega n_{0}\delta v_{ez}
=\frac{q_{e}}{m}\Gamma_{0}\delta E_{z}-\frac{q_{e}}{mc^{2}}U_{t}^{2} n_{0}\delta E_{z}, \end{equation}
and
\begin{equation}\label{RHD2018Cl eq E z lin with B} (\omega^{2}-k^{2}c^{2})\delta E_{z}+4\pi\imath\omega q_{e}n_{0e}\delta v_{ez}=0 \end{equation}
Equations (\ref{RHD2018Cl eq v z lin with B}) and (\ref{RHD2018Cl eq E z lin with B}) give the following spectrum:
\begin{equation}\label{RHD2018Cl spectrum transverse} \omega^{2}=k^{2}c^2+\omega_{Le}^{2}\biggl(\frac{K_{1}}{K_{2}}-\frac{U_{t}^{2}}{c^{2}}\biggr) \end{equation}

It gives same result as the electromagnetic one-dimensional waves in the relativistic isotropic plasmas.

Equation (\ref{RHD2018Cl spectrum transverse}) shows that
the transverse waves includes same reduction of the Langmuir frequency
as it happens for the longitudinal waves (\ref{RHD2018Cl Langmuir wave spectrum}).

\section{Conclusion}

Hydrodynamic model for the relativistically hot ideal plasmas has been developed as a chain of equations.
Derivation of equations is performed for functions
which appear in previous equations.
For example, derivation of the continuity equation leads to the current of particles.
Next, the equation for evolution of the current has been derived.
The particle current evolution contains four novel functions.
Two of them are the three-scalar and three-vector.
Therefore, equations for the evolution of these functions have been found.
So, no functions is introduced \textit{ad hoc}.
Therefore, no energy-momentum tensor is involved.
In stead of it two four-tensors relatively to equations for two four-vectors have been found.
They are the four-particle current,
which also exists in the Maxwell equations as the source of field,
and the four-Gamma vector,
which is the average of the reverse gamma factor,
and the average of the thermal velocity divided by the relativistic gamma factor.
These variables are chosen for a simple reason.
The evolution of four-particle current leads to appearance of four-Gamma vector.
While the evolution of four-Gamma vector produces the four-particle current.
Therefore, the set of equations tries to close itself.
Few new functions similar to the pressure appear either.
Therefore, the set of equations is not completely closed.
Consequently, the chain of equations can be continued.
Or, the chain can be truncated at this stage forming an example of minimal coupling model for relativistic plasmas.

Relativistic plasmas show particular interest in the regime of the relativistically hot plasmas.
Consequently, details of evolution in the momentum space are essential.
However, the application of kinetic equations can be rather complicate.
Therefore, we need as much of higher moments as it is possible to cover the reach kinetic behavior.

Presented model does not cover high tensor dimensional hydrodynamic functions,
but consider two variations of four-vectors.
There is the third variation of four-vector.
It is the four-momentum.
However, it is not appear in obtained equations.

Derivation of hydrodynamic equations starts with the definition of concentration,
where the integral operator explicitly averages the microscopic dynamics over the macroscopically small volume.
The truncation presiger is discussed after derivation of basic equations.

The closed set of equations is applied for study of fundamental collective phenomena in relativistic plasmas.
The spectrum of Langmuir waves is obtained.
The dispersion equation for the cold relativistic beam propagating through macroscopically motionless relativistically hot plasmas is demonstrated.
The cold regimes, particularly, the beam propagation, shows agrement with the well-known models of relativistic plasmas motion.
While, the thermal effects, particularly in the magnetic field, gives new information about plasma dynamics in hydrodynamic regime
(these effects are discussed in earlier papers).

\section{Acknowledgements}
Work is supported by the Russian Foundation for Basic Research (grant no. 20-02-00476).
This paper has been supported by the RUDN University Strategic Academic Leadership Program.

\section{DATA AVAILABILITY}

Data sharing is not applicable to this article as no new data were
created or analyzed in this study, which is a purely theoretical one.



\end{document}